\documentstyle[a4,12pt]{article}
\renewcommand{\baselinestretch}{1.2}
\begin{document}
\def\dx{\partial_x}
\def\deltamn{\delta_{m+n,0}}
\def\deltaxy{\delta(x-y)}
\def\levi{\epsilon_{ijk}}

%\topmargin 0pt
%\advance \topmargin by -\headheight
%\advance \topmargin by -\headsep
%\marginparwidth 0.5in
%               macros formatting and equations
%%       fake blackboard bold macros for reals, complex, etc.
\def\rlx{\relax\leavevmode}
\def\inbar{\vrule height1.5ex width.4pt depth0pt}
\def\IZ{\rlx\hbox{\sf Z\kern-.4em Z}}
\def\IR{\rlx\hbox{\rm I\kern-.18em R}}
\def\ID{\rlx\hbox{\rm I\kern-.18em D}}
\def\IC{\rlx\hbox{\,$\inbar\kern-.3em{\rm C}$}}
\def\one{\hbox{{1}\kern-.25em\hbox{l}}}

\def\eF{\cal F}
\def\eG{\cal G}
\def\eD{\cal D}
\def\eH{\cal H}
\def\eP{\cal P}
\def\eQ{\cal Q}
\def\eO{\cal O}
\def\eE{\cal E}
\def\smallfrac#1#2{\mbox{\small $\frac{#1}{#2}$}}

%
% Uncomment this block to use euler fonts
%
%\font\teneusm=eusm10 scaled 1200
%\font\seveneusm=eusm7 scaled 1200
%\font\fiveeusm=eusm5 scaled 1200
%\newfam\eusmfam
%\textfont\eusmfam=\teneusm
%\scriptfont\eusmfam=\seveneusm
%\scriptscriptfont\eusmfam=\fiveeusm
%\def\eusm#1{{\fam\eusmfam\relax#1}}
%\def\eF{\eusm F}
%\def\eG{\eusm G}
%\def\eD{\eusm D}
%\def\eH{\eusm H}
%\def\eP{\eusm P}
%\def\eQ{\eusm Q}
%\def\eO{\eusm O}
%\def\eE{\eusm E}
%

\begin{titlepage}

October, 1992 \hfill{UTAS-PHYS-92-32}

\hfill{hep-th/9301077}
\vskip 1.6in
\begin{center}
{\Large {\bf The $N=2$ super $W_4$ algebra and its }}\\[10pt]
{\Large{\bf associated generalized KdV hierarchies}}
\end{center}

\normalsize
\vskip .4in

\begin{center}
C. M. Yung  \hspace{3pt}
and \hspace{3pt} Roland C. Warner
\par \vskip .1in \noindent
{\it Department of Physics, University of Tasmania}\\
{\it GPO Box 252C Hobart, Australia 7001}
\end{center}
%\footnotetext{ }
\par \vskip .3in

\begin{center}
{\Large {\bf Abstract}}\\
\end{center}
\par \vskip .3in \noindent
We construct the $N=2$ super $W_4$ algebra as a certain reduction of
the second Gel'fand-Dikii
bracket on the dual of the Lie superalgebra of $N=1$ super
pseudo-differential operators.
The
algebra is put in manifestly $N=2$ supersymmetric form in terms of three $N=2$
superfields $\Phi_i(X)$, with $\Phi_1$ being the $N=2$ energy momentum
tensor and $\Phi_2$ and $\Phi_3$ being conformal spin $2$ and $3$
superfields respectively. A search for integrable hierarchies of the
generalized KdV variety with this algebra as Hamiltonian structure
gives three solutions, exactly the same number as for the $W_2$
(super KdV) and $W_3$ (super Boussinesq) cases.
\end{titlepage}

\vspace{1cm}

\section{Introduction}

The study of conformal field theories has been important to string
theory \cite{gsw},
critical statistical mechanics in two dimensions \cite{bpz} and two
dimensional quantum gravity \cite{qg}. The role of the Virasoro
algebra and its extensions has been crucial in this study. Of particular
interest in recent times is the the class of
extended conformal algebras known as $W$ algebras. Not long after its
introduction to conformal field theory \cite{zamo}, it was realized
\cite{walg} that its classical version had been studied earlier in the
context of integrable systems \cite{gd,adler}.  What have since become known
as the classical $W_n$ algebras turn out to provide the second Hamiltonian
structure of the generalized KdV hierarchies.

Knowledge of the generalized KdV hierarchies can be expected to give
insight into the structure of the $W$ algebras. An early work in this
direction is \cite{fateev} in which essential use is made of the Miura
transformation to free fields. The quantization of the generalized KdV
hierarchies \cite{quantization}, which can be regarded as integrable
classical field theories, provides information on perturbations of
conformal field theories \cite{zamo2} and, in turn, off-critical
statistical mechanics. The generalized KdV hierarchies are themselves
of major interest in the matrix models of quantum gravity \cite{qg}.

The situation with supersymmetric $W$ algebras and supersymmetric
generalized KdV hierarchies is not as clear at present. It is
probably fair to say that the $N=2$ theories are currently better understood.
These
are also of particular interest to string theory, and will be our
sole concern here. It is known
that there exist three $N=2$ supersymmetric KdV hierarchies
\cite{laberge,labelle}. The construction
of $N=2$ classical super $W$ algebras was recently proposed
\cite{figueroa,inami,huitu}.
Three supersymmetric Boussinesq equations were found
\cite{yung} to be associated with the $N=2$ super $W_3$ algebra.  In this
paper we construct the $N=2$ super $W_4$ algebra and argue for the
existence of three integrable supersymmetric
hierarchies with it as Hamiltonian
structure. Based on this and on \cite{labelle} and \cite{yung} we
conjecture that there are three generalized supersymmetric KdV hierarchies
associated with each $N=2$ super $W_n$ algebra.

\section{The classical $N=2$ super-$W_4$ algebra}

The construction of the classical
$N=2$ super $W_n$ algebras in terms of $N=1$ superfields was
proposed independently in \cite{figueroa}, \cite{inami} and
\cite{huitu}. The $N=2$ super $W_3$ was
explicitly worked out in the latter two papers.

Let us first make our notation clear.  We will be working in both
$(1|1)$ and $(1|2)$ superspace. Consider thus
the superspace $(1|N)$ with coordinates
$X=(x,\theta_1,\ldots,\theta_N)$, and covariant
superderivative $D_i=\partial /\partial\theta_i + \theta_i\partial$ obeying
$D_i^2=\partial\equiv\partial_x$ and $D_iD_j=-D_jD_i$ for $i\neq j$.
The Berezin integral of a super function $f(X)$ is defined as
$\int_B f(X) \equiv \int f(X) dX$ with $dX=d\theta_N\cdots d\theta_1 dx$.
We use the convention
$\int \theta_i d\theta_j  = \delta_{ij}$ and $\int d\theta_j =0$.
The Grassmann parity of a super function $f(X)$ is denoted by $|f|$,
and takes the value $0$ (resp. $1$) if $f$ is even (resp. odd).

We next introduce several
objects required to do variational calculus in superspace.
Given a super differential operator (SDO)
$L$, we define its adjoint $L^*$ with
respect to the inner product $(f,g)\equiv \int_B fg$ through the relation
\begin{displaymath}
(Lf,g)=(-1)^{|L||f|}(f,L^* g),
\end{displaymath}
which consquently has the properties
(i) $(L^*)^*=L$ and (ii) $(PQ)^* = (-1)^{|P||Q|} Q^* P^*.$
Given a functional ${\eF}[\Phi] = \int_B f[\Phi]$
of the superfields $\Phi(X)=(\Phi_1(X),\Phi_2(X),\ldots)$,
the variational derivative $\delta {\eF}/
\delta \Phi_j$ is defined through the relation
\begin{displaymath}
\int_B \Gamma \frac{\delta {\eF}}{\delta \Phi_j} =  \left.
	  \frac{d}{d\epsilon}\right|_{\epsilon=0}
	  {\eF}[\ldots,\Phi_j+\epsilon \Gamma,\ldots].
\end{displaymath}
In $N=1$ superspace, it is given explicitly by
\begin{displaymath}
  \frac{\delta {\eF}}{\delta \Phi_j} = \sum_{k=0}^{\infty}
 (-1)^{|\Phi_j|k + k(k+1)/2}
  D^k \frac{\partial f}{\partial (D^k\Phi_j)},
\end{displaymath}
whereas in $N=2$ superspace the corresponding expression is
\begin{eqnarray}
  \frac{\delta {\eF}}{\delta \Phi_j} & = & \sum_{k=0}^{\infty}
   (-1)^k \partial^k \left(
    \frac{\partial}{\partial (\partial^k\Phi_j)} - (-1)^{|\Phi_j|}
   \sum_{i=1}^{2}
     D_i \frac{\partial}{\partial (\partial^kD_i\Phi_j)} \right.\nonumber\\
     & & + \left.
    D_1D_2 \frac{\partial}{\partial (\partial^kD_1D_2\Phi_j)}\right)f.
    \nonumber
\end{eqnarray}

It was proposed in \cite{figueroa,inami,huitu} that the
construction of the $N=2$ super $W_n$ algebra be performed in
$(1|1)$ superspace.
Given a set of $N=1$ superfields $u_i(X)$, let ${\cal G}$ be the infinite
dimensional Lie superalgebra of super pseudo-differential
operators ($S\Psi$DO's), which are formal Laurent series in
$D^{-1}\equiv D\partial^{-1}$
with coefficients which are differential polynomials in $u_i(X)$.
Multiplication in the algebra is
defined through the usual Leibniz rule for superdifferentiation
augmented by
the rules $\partial \partial^{-1}=
\partial^{-1}\partial=1$ and
\begin{displaymath}
   \partial^kf =f\partial^k + \sum_{i=1}^{\infty}\left(\begin{array}{c}
   k\\i\end{array}\right)(\partial^if)\partial^{k-i},
\end{displaymath}
for any integer $k$.
With the super residue $sres P$ of a
S$\Psi$DO $P=\sum p_i D^i$ being $p_{-1}$, the Adler super trace
$Str AB \equiv
\int_B  sres AB$ is well defined \cite{manin}, with the property
$Str(AB)=(-1)^{|A||B|}Str(BA)$.
The Adler super trace defines a nondegenerate supersymmetric bilinear
invariant form $\langle , \rangle$ on ${\cal G}$ via
\begin{displaymath}
\langle A,B \rangle = Str(AB),
\end{displaymath}
allowing the identification of ${\cal G}$ with its dual ${\cal G}^*$.

The Lie superalgebra ${\cal G}$ splits into a direct sum of the
subalgebra ${\cal G}_+$ of SDO's and the subalgebra
${\cal G}_-$ of ``integrational operators'', with ${\cal G}_{\pm}$
being dual to ${\cal G}_{\mp}$. The projection of an element
$Y\in {\cal G}$ into ${\cal G}_{\pm}$ is denoted by $Y_{\pm}$. We
concentrate on the subspace ${\cal G}_n$ of (homogeneous Grassmann
parity) SDO's of
the form
\begin{equation}
L=D^n + u_{n-1}D^{n-1} + \cdots + u_0.
\label{eqn:L}
\end{equation}
It was shown in \cite{fig1} that the map $J : {\cal G}_n^* \rightarrow
{\cal G}_n$ defined by
\begin{equation}
J(Y)=L(YL)_+-(LY)_+L
\end{equation}
defines a Poisson
bracket (the ``second'' Gel'fand-Dikii bracket)
on the space of functionals of $u_i(X)$ through
\begin{equation}
\{{\eF},{\eG}\}=\langle J(d{\eF}),d{\eG}\rangle,
\label{eqn:gdbracket}
\end{equation}
with $d{\eF}$ being the gradient of ${\eF}$ with respect to the
bilinear form $\langle , \rangle$:
\begin{displaymath}
\langle A,d{\eF}\rangle=\left.\frac{d}{d\epsilon}\right|_{\epsilon=0}
{\eF}[L+\epsilon A].
\end{displaymath}
An analogous Poisson bracket structure
can in fact be associated to any Lie superalgebra admitting
a trace form and a ``unitary Yang-Baxter operator $R$''. For details see
\cite{yungyb}.
``Coordinatizing'' $L$ and $A$ by (\ref{eqn:L}) and
$A={\sum}_{k=1}^{n-1}A_k D^k$
respectively,
$d{\eF}$ can be shown to be given by
\begin{equation}
d{\eF}= (-1)^{|{\eF}|+|L|+1}
     \sum_{k=0}^{n-1}(-1)^kD^{-k-1}\frac{\delta {\eF}}{\delta u_k}.
\label{eqn:df}
\end{equation}

The $N=2$ super $W_m$ algebra is the Poisson bracket algebra
(\ref{eqn:gdbracket}) induced on the subspace ${\cal G}_{2m-1}^{(0)}$
of ${\cal G}_{2m-1}$ of SDO's with vanishing
coefficient of $D^{2m-2}$. Actually, what has been proved so far
\cite{figueroa} is
that this reduction contains the $N=2$ super Virasoro algebra
as a subalgebra. It remains a conjecture that $(m-2)$
other $N=2$ primary fields can be constructed out of the
$u_i(X)$, and that the Poisson brackets amongst the primary fields close
on themselves and their derivatives.
This was shown \cite{inami,huitu} to be true for $m=3$. We will show
that it is also true for $m=4$. To obtain the Poisson bracket on the subspace
${\cal G}_{2m-1}^{(0)}$, one can either use the Dirac procedure or,
equivalently
\cite{figueroa}, the following: define the gradient $d{\eF}$ of
a functional on ${\cal G}_{2m-1}^{(0)}$
as (\ref{eqn:df}) (with $n=2m-1$) but with $\delta {\eF}/
\delta u_{2m}$ replaced by $Y_{2m-2}$ such that $sres[L,d{\eF}]=0$.

We now concentrate on the construction of the $m=4$ case.
The two $N=1$ superfields which generate the classical $N=2$ Virasoro
subalgebra of $N=2$ super $W_4$
are given by \cite{figueroa}
\begin{eqnarray}
J(X) & = & u_{5},\\
T(X) & = & u_{4} - \frac{1}{2}(Du_{5}).
\end{eqnarray}
According to the general theory \cite{figueroa},
four $N=1$ primary fields $W_k(X)$ of
conformal spins $k=2$, $\frac{5}{2}$, $3$ and $\frac{7}{2}$ organized into the
$N=2$ supermultiplets
$(W_{2}(X),W_{5/2}(X))$ and $(W_{3}(X),W_{7/2}(X))$ can be constructed out of
the $u_i(X)$ and their derivatives. To determine $W_k(X)$, we look for the
most general combination of the $u_i(X)$ of spin $k$ (or ``degree'' $k$ - the
degree of $\partial$ and $D$ being $1$ and $1/2$ respectively) such that its
Poisson bracket with $T(X)$ is given by
\begin{equation}
\{W_k(X),T(X')\} = \left( (\partial W_k(X)) + kW_k(X)\partial - (-1)^{|W_k|}
\frac{1}{2}(DW_k(X))D\right)\Delta(X-X'),\label{WkTPoissonBracket}
\end{equation}
where $\Delta(X-X')=(\theta-\theta')\delta(x-x')$ is the $N=1$ supersymmetric
delta function.
The Poisson bracket is calculated from (\ref{eqn:gdbracket}) and, although
straightforward, is long and tedious. A program in
Mathematica
\cite{mathematica} was written to handle the calculation. To check that the
program gives the desired result, the following alternative way to
determine $W_k(X)$ was also employed.

Primary fields of a given conformal spin can also be constructed by the
requirement that
the $SDO$ (\ref{eqn:L}) (with $u_{n-1}=0)$ is a super covariant operator.
That is, we require $L$ to transform as
\begin{equation}
L= (D\tilde{\theta})^{\frac{n+1}{2}}\tilde{L}(D\tilde{\theta})^{\frac{n-1}{2}}.
\end{equation}
under superconformal transformations \cite{Friedan}
$X \rightarrow
\tilde{X}(X)=\{\tilde{x}(x,\theta),\tilde{\theta}(x,\theta)\}$ (with
$D=(D\tilde{\theta})\tilde{D}$). Primary fields of conformal spin $k$
transform under such superconformal transformations
as $W_k=\widetilde{W_k}(D\tilde{\theta})^{2k}$.

First
the transformation properties of the $u_i$ fields are determined from
those of $L$.
Substituting $(\tilde{D})^j=((D\tilde{\theta})^{-1}D)^j$ into $\tilde{L}$,
expanding and equating coefficients of $D$ yields the superconformal
transformation properties of the $u_i$ functions
\begin{eqnarray}
u_5& = &\tilde{u_5}(D\tilde{\theta})^2\nonumber\\
u_4& = &\tilde{u_4}(D\tilde{\theta})^3
+\tilde{u_5}(D\tilde{\theta})(D^2\tilde{\theta}) +6S\nonumber\\
u_3& = &\tilde{u_3}(D\tilde{\theta})^4
- 2 \tilde{u_4}(D\tilde{\theta})^2(D^2\tilde{\theta})
+2 \tilde{u_5}(D\tilde{\theta})(D^3\tilde{\theta}) +4DS\nonumber\\
u_2& = &\tilde{u_2}(D\tilde{\theta})^5 +2\tilde{u_3}(D\tilde{\theta})^3(D^2
\tilde{\theta}) +4\tilde{u_4}(D\tilde{\theta})^2(D^3\tilde{\theta})\nonumber\\
& + &\tilde{u_5}\{5(D\tilde{\theta})(D^4\tilde{\theta})
-6(D^2\tilde{\theta})(D^3\tilde{\theta})\}+
8D^2S\nonumber\\
u_1& = &\tilde{u_1}(D\tilde{\theta})^6- \tilde{u_2}(D\tilde{\theta})^4(D^2
\tilde{\theta})+2\tilde{u_3}(D\tilde{\theta})^3(D^3\tilde{\theta})\nonumber\\
& - &\tilde{u_4}\{(D\tilde{\theta})^2(D^4\tilde{\theta})+2(D\tilde{\theta})
(D^2\tilde{\theta})(D^3\tilde{\theta})\}  \nonumber\\
& + &\tilde{u_5}\{(D^2\tilde{\theta})(D^4\tilde{\theta})+2(D\tilde{\theta})(D^5
\tilde{\theta})-2(D^3\tilde{\theta})^2\}+2D^3S\nonumber\\
u_0& = &\tilde{u_0}(D\tilde{\theta})^7
+3\tilde{u_1}(D\tilde{\theta})^5(D^2\tilde{\theta})
+3\tilde{u_2}(D\tilde{\theta})^4(D^3\tilde{\theta}) \nonumber\\
& + &3\tilde{u_3}(D\tilde{\theta})^3(D^4\tilde{\theta}) +3\tilde{u_4}\{
(D\tilde{\theta})^2(D^5\tilde{\theta})-(D\tilde{\theta})(D^2\tilde{\theta})
(D^4\tilde{\theta})\}\nonumber\\
& + &3 \tilde{u_5}\{(D\tilde{\theta})(D^6\tilde{\theta})-(D^3\tilde{\theta})
(D^4\tilde{\theta})-(D^2\tilde{\theta})(D^5\tilde{\theta})\} +3D^4S +9SDS.
\label{superconfrules}
\end{eqnarray}
The first term in each equation is the appropriate transformation for a primary
field of $1$, $\frac{3}{2}$, $2$, $\frac{5}{2}$, $3$, $\frac{7}{2}$ while the
remaining terms
involve more complicated terms of the same degree showing that the $u_i$ are
not
primary fields, and also include the
super-Schwartzian of the superanalytic map
\begin{equation}
S=\frac{D^4\tilde{\theta}}{D\tilde{\theta}} -
2 \frac{(D^3\tilde{\theta})
(D^2\tilde{\theta})}{(D\tilde{\theta})^2},
\end{equation}
and its superderivatives. The super-Schwartzian terms signify
the presence of a central term in the corresponding operator
product expansion and in fact they prove essential in enabling
the construction of primary fields for the higher conformal spins. We consider
the most general combinations of the $u_i$'s and their derivatives consistent
with the desired degree, including products of $u_i$'s, and using the
explicit transformation properties (\ref{superconfrules}) we determine the
coefficients required to obtain  primary fields. This method, which
was used in \cite{huitu} for the case of $N=2$ super $W_3$, confirms the
form of the primary fields deduced by implementing the superconformal
transformations using the Poisson bracket (\ref{WkTPoissonBracket}).

The primary fields $W_k(X)$ were found to be
\begin{eqnarray}
W_{2}(X) & = & u_3 -\smallfrac{2}{3}(Du_4) -\smallfrac{2}{3}(\partial u_5)+
	   \alpha(u_5)^2,\\
W_{5/2}(X) & = & u_2 - \smallfrac{1}{2}(Du_3) - (\partial u_4) +
	  \smallfrac{1}{3}(\partial D u_5) - \smallfrac{11}{18} u_5 T\\
W_{3}(X) & = & u_1 - \smallfrac{1}{5}(Du_2) -\smallfrac{2}{5}(\partial u_3)
   + \smallfrac{1}{5}(\partial Du_4) +\smallfrac{1}{10}(\partial^2u_5)
     +\beta u_5u_3 \nonumber\\
& + &\gamma u_5^3 + \smallfrac{3}{20}(Du_5)u_4 +
(\smallfrac{1}{20}-\smallfrac{2}{3}\beta)u_5(\partial u_5) -
(\smallfrac{1}{10}+\smallfrac{2}{3}\gamma)u_5(Du_4)\\
W_{7/2}(X) & = & u_0 - \smallfrac{1}{2}(Du_1) -\smallfrac{1}{2}(\partial u_2)
+ \smallfrac{1}{5}(\partial D u_3) + \smallfrac{1}{5}(\partial^2u_4)\nonumber\\
& - & \smallfrac{1}{20}(\partial^2D u_5) + \delta u_5(Du_3) + \epsilon u_5^2u_4
+ \lambda u_3(Du_5) -\smallfrac{1}{20}(Du_4)u_4 \nonumber\\
& - & \smallfrac{3}{10}u_3u_4 + \smallfrac{13}{40}(\partial u_5)u_4 -
\smallfrac{1}{2}\epsilon
u_5^2(Du_5) +  \smallfrac{1}{20}\left( 3-20\lambda - 40\delta\right)u_5u_2
\nonumber\\
& + &\smallfrac{1}{15}\left( -3 + 20 \lambda + 30\delta\right) u_5(\partial
u_4)
\nonumber\\
& + &\smallfrac{1}{120}\left(-33+ 220\lambda - 360\epsilon + 360\delta\right)
u_5(\partial D u_5)\nonumber\\
& + &\smallfrac{1}{60}\left(21-205\lambda + 270\epsilon - 330\delta\right)
(\partial u_5)(Du_5)
\end{eqnarray}
Note the presence of six undetermined constants in the above.
In order that $(W_2(X),W_{5/2}(X))$ and $(W_3(X),W_{7/2}(X))$ form $N=2$
supermultiplets, we require $W_2(X)$ and $W_3(X)$ to have the following
Poisson brackets with the $U(1)$ current $J(X)$:
\begin{eqnarray}
\{W_2(X),J(X')\} & = & - 2 W_{5/2} \Delta(X-X')\nonumber\\
\{W_3(X),J(X')\} & = & - 2 W_{7/2} \Delta(X-X').\nonumber
\end{eqnarray}
Together with the corresponding brackets between $W_{5/2}(X)$ and $J(X)$
and $W_{7/2}(X)$ and $J(X)$ implied by the above and by the Jacobi identity,
we find that the constants $\alpha$ - $\lambda$ are fixed {\em uniquely} :
\begin{displaymath}
\alpha=-\frac{11}{36}, \hspace{6pt} \beta=-\frac{3}{10}, \hspace{6pt}
\gamma=\frac{23}{360}, \hspace{6pt} \delta=\frac{3}{20} , \hspace{6pt}
\epsilon = \frac{23}{160}, \hspace{6pt} \lambda = \frac{3}{20}
\end{displaymath}
In accordance with expectations, we find that the Poisson brackets amongst the
primary fields $W_k(X)$ with these values of $\alpha$ - $\lambda$ close
on themselves and their derivatives. We can thus say that we have
constructed the classical $N=2$ super $W_4$ algebra.

There is no point in displaying the Poisson brackets amongst the remaining
$N=1$ primary superfields.
Apart from the fact that it would take more than 10 pages, there is a more
natural and
compact formalism - namely the $N=2$ formalism. In fact it would be preferable
for the algebra to be constructed in $N=2$ superspace from scratch, presumably
as a reduction of the Gel'fand-Dikii algebra on the dual of the Lie algebra
of $N=2$ S$\Psi$DO's. How to carry this out is however an open problem.
Let us define the $N=2$ superfields
\begin{eqnarray}
\Phi_1(x,\theta_1,\theta_2) & = & \alpha_1 \theta_1 T(x,\theta_2) +
   \alpha_2 J(x,\theta_2),\\
\Phi_2(x,\theta_1,\theta_2) & =  & \beta_1 \theta_1 W_{5/2}(x,\theta_2) +
   \beta_2 W_2(\theta_2,x), \\
\Phi_3(x,\theta_1,\theta_2) & = & \gamma_1 \theta_1 W_{7/2}(x,\theta_2) +
   \gamma_2 W_3(\theta_2,x)
\label{eqn:phi}
\end{eqnarray}
with conformal dimension $1$, $2$ and $3$ respectively. Without loss of
generality we choose $\alpha_1=1$. We find that the
bracket $\{ \Phi_1(X), \Phi_1(X')\}$  closes on $\Phi_1$ and its
derivatives if and only if $\alpha_2 = \pm \frac{i}{2}$. We choose the
positive sign. We then find that $\{\Phi_2(X),\Phi_1(X')\}$ and
$\{\Phi_3(X),\Phi_1(X')\}$ close on $\Phi_i$ and derivatives if and only
if $\beta_2=\frac{i}{2}\beta_1$ and $\gamma_2=\frac{i}{2}\gamma_1$. In that
case, all the other brackets close on $\Phi_i$ and derivatives as well. We
find the choice $\beta_1=i$ and $\gamma_1=1$ convenient. In this case, all
the coefficients on the right hand side of the brackets are real.

The Poisson brackets amongst the $\Phi_i$ are given by
\begin{eqnarray}
\{\Phi_1(X),\Phi_1(X')\} & = & \left(3\partial D_1 D_2 -
    \Phi_1 \partial + \smallfrac{1}{2}(D_i\Phi_1)D_i - (\partial\Phi_1)\right)
    \Delta(X-X') \label{eqn:pbuu}\\
\{\Phi_2(X),\Phi_1(X')\} & = & \left(-2\Phi_2\partial +
\smallfrac{1}{2}(D_i\Phi_2)D_i -
    (\partial\Phi_2)\right)\Delta(X-X')\label{eqn:pbbu}\\
\{\Phi_3(X),\Phi_1(X')\} & = & \left(-3\Phi_3\partial +
\smallfrac{1}{2}(D_i\Phi_3)D_i -
    (\partial\Phi_3)\right)\Delta(X-X')\label{eqn:pbcu}\\
\{\Phi_2(X),\Phi_2(X')\} & = & \left( {\eO}_2 - {\eO}_2^*\right)
\Delta(X-X') \label{eqn:pbbb}\\
\{\Phi_3(X),\Phi_2(X')\} & = & {\eD}_{32} \Delta(X-X') \label{eqn:pbcb}\\
\{\Phi_3(X),\Phi_3(X')\} & = & \left( {\eO}_3 - {\eO}_3^*\right)
\Delta(X-X') \label{eqn:pbcc},
\end{eqnarray}
where
$\Delta(X-X')\equiv (\theta_1-\theta_1')(\theta_2-\theta_2')
\delta(x-x')$
is the $N=2$ supersymmetric delta function and ${\eO}_2$, ${\eO}_3$
and ${\eD}_{32}$ are displayed in the Appendix.
The fields on the right-hand sides of (\ref{eqn:pbuu}) - (\ref{eqn:pbbb})
are evaluated at the point $X$.
The
Poisson bracket (\ref{eqn:pbuu}) defines the classical $N=2$ super Virasoro
algebra. The brackets (\ref{eqn:pbbu}) and (\ref{eqn:pbcu})
simply state that $\Phi_k$ is an
$N=2$ primary field with conformal dimension $k$, for $k=2,3$.
We summarize the Poisson brackets (\ref{eqn:pbuu}) - (\ref{eqn:pbcc})
amongst the primary fields $\Phi_k$ in the form
\begin{equation}
   \{\Phi_i(X),\Phi_j(X')\}={\eD}_{ij}\Delta(X-X').
\label{eqn:pb}
\end{equation}
The matrix-valued Hamiltonian operator ${\eD}$ is anti-selfadjoint:
${\eD}_{ij}^{*}=-{\eD}_{ji}$, reflecting the anti-(super)symmetry of the
Poisson brackets.
The Poisson bracket  can be
extended to arbitrary functionals ${\eF}[\Phi]$ and ${\eG}[\Phi]$ via
\begin{equation}
  \left\{{\eF},{\eG} \right\}= \int_B \left({\eD}_{ij}\frac{\delta {\eF}}
  {\delta
  \Phi_j}
  \right)
   \frac{\delta {\eG}}{\delta \Phi_i}. \label{eqn:n2pb}
\end{equation}

\section{The associated integrable hierarchies}

Gel'fand-Dikii brackets are intimately connected with integrable Hamiltonian
systems. In fact they first arose out of a study of generalized KdV hierarchies
\cite{gd}. One therefore expects that the $N=2$ super $W_4$ algebra constructed
in the last section can also be constructed out of a Lax operator for an
integrable hierarchy. A first step towards making such a connection concrete
is to look for the integrable hierarchy (or hierarchies) associated
with the algebra - namely an infinite
set of Hamiltonian functionals mutually in involution with respect to the
Poisson bracket (\ref{eqn:pb}).
Our approach is constructive and follows that of
\cite{labelle}. We will show the existence of three families of Hamiltonian
functionals in involution, assuming only space-translation invariance.
Naturally
the extent of these families as constructed is finite, but we conjecture that
they constitute the first few members of each of three infinite families of
Hamiltonian functionals in involution.

Let us first note that there is an
$S_2$ symmetry in the Poisson bracket algebra, which is also present in the
$N=2$ super $W_2$ (Virasoro) case \cite{laberge} and in the $N=2$ super
$W_3$ case \cite{yung}.
For the permutation $(12)$ define the map
$\Pi_{(12)}$ acting on the superfields by $\Pi_{(12)}:\Phi_1 \mapsto -\Phi_1,
\Phi_2 \mapsto \Phi_2, \Phi_3 \mapsto -\Phi_3,
D_1 \mapsto D_2, D_2 \mapsto D_1,  dX \mapsto - dX$.
Then $\Pi_{(12)}$ together with the
identity map $\Pi_e$ forms a representation of the symmetric group $S_2$.
We say that a superfield
$f$ is $\Pi$-even if $\Pi_{(12)} f = f$, $\Pi$-odd if $\Pi_{(12)} f = -f$ and
$\Pi$-definite if it is either $\Pi$-even or $\Pi$-odd.
The symmetric or $\Pi$-even
(anti-symmetric or $\Pi$-odd) part of a superfield $f$ is given by $f_s=Pf$
($f_a=Qf$) where
\begin{eqnarray}
P &=& \sum_{x\in S_2}\Pi_x\nonumber\\
Q &=& \sum_{x\in S_2}\delta_x \Pi_x\nonumber
\end{eqnarray}
with $\delta_x$ being the parity of the permutation $x$, are respectively the
symmetrizer and anti-symmetrizer for the group. One can check that the Poisson
bracket (\ref{eqn:pb}) respects the $S_2$ symmetry. By this we mean the
following: Firstly, note that $\Pi_{(12)}(\delta/\delta \Phi_j)=(-1)^j
\delta/\delta \Phi_j$ for $j=1,2,3$. Note also that $\Pi_{(12)}({\eD}_{ij})
=-(-1)^{i+j}{\eD}_{ij}$. Therefore we have $\Pi_{(12)} \{{\eF},{\eG}\}=
\{ \Pi_{(12)}({\eF}), \Pi_{(12)}({\eG}) \}$. In particular, this means that
if $\{{\eF},{\eG}\}=0$ and that ${\eF}$ is $\Pi$-definite, then ${\eF}$ is in
involution with both $Q({\eG})$ and $P({\eG})$. A similar argument holds for
the degree - if one of the Hamiltonian functionals
of the hierarchy is of homogeneous
degree, then all other Hamiltonians can be taken to be of homogeneous degree.

We now note the importance of the Hamiltonian functional ${\eH}_1^{+}=\int_B
\Phi_1$. Its Poisson bracket with any Hamiltonian functional ${\eH}_k[\Phi]=
\int_B h_k[\Phi]$,
for which there is no explicit $X$ dependence, is
given by
\begin{eqnarray}
\{ {\eH}_1^{+}, {\eH}_k \} & = & \int_B \left({\eD}_{i1}(1)\right)\frac{\delta
    {\eH}_k}{\delta \Phi_i}\nonumber\\
   & = & -\int_B(\partial \Phi_i) \frac{\delta {\eH}_k}{\delta \Phi_i}
\nonumber\\
   & = & - \left.\frac{d}{d\epsilon}\right|_{\epsilon=0} {\eH}_{k}
      [\Phi(x+\epsilon)] \nonumber\\
   & = & 0.\nonumber
\end{eqnarray}
Put another way, the Hamiltonian vector field corresponding to ${\eH}_1$ is
the generator of translations in $x$, and as long as we require the flows
generated by the ${\eH}_k$ which form
the hierarchy to be invariant under $x$-translations, as is the case for
the generalized KdV hierarchies, ${\eH}_1^{+}$ must belong to the hierarchy.
Since ${\eH}^{+}_1$ is $\Pi$-definite and of homogeneous degree,
we have the result that all the members
${\eH}_k$ are also $\Pi$-definite and of homogeneous degree.

To write down the most general $\Pi$-even (resp. odd)
${\eH}_n=\int_B h_n $ of a particular degree, we construct
all the independent $\Pi$-odd (resp. even)
densities $h_n$. Two densities $f$ and $g$
are dependent if there exists non-trivial $\alpha$ and $\beta$ such that
$\int_B ( \alpha f + \beta g ) =0$ or, equivalently, if $\alpha \frac{\delta}
{\delta \Phi_j}\int_B f + \beta \frac{\delta}{\delta \Phi_j} \int_B g  =0$
for all $j$.
A program written in Mathematica \cite{mathematica} was used to generate
the ${\eH}^{\pm}_n$'s, where the subscripts indicate the degrees and
superscripts the $\Pi$-parities.

To determine when two Hamiltonian functionals are in involution, we need a
further tool. We define the Fr\'{e}chet derivative $\ID_K = (\ID_K^1,
\ID_K^2,\ldots)$ of a functional $K[\Phi]$ to be such that its component
$\ID^j_K$ is given by
\begin{displaymath}
\ID^j_K(\Gamma)=(-1)^{|\Gamma|(|K|+|\Phi_j|)}
\left.\frac{d}{d\epsilon}\right|_{\epsilon=0}
K[\ldots,\Phi_j+\epsilon \Gamma,\ldots],
\end{displaymath}
for any $\Gamma[\Phi]$.
The Fr\'{e}chet derivative has the properties
\begin{eqnarray}
\ID^j_{(\partial^m D_1^nD_2^p
\Phi_j)}& = & (-1)^{|\Phi_j|(n+p)}\partial^m D_1^n D_2^p\nonumber\\
   \ID_{fg}^j & = &  (-1)^{|f||\Phi_j|}f\ID_g + (-1)^{|g|(|\Phi_j|+|f|)}g\ID_f.
\nonumber
\end{eqnarray}
In fact it has an intimate connection with the variational derivative.
Namely, one can show that if ${\eF}=\int_B f$ then
\begin{eqnarray}
\frac{\delta {\eF}}{\delta \Phi_j} & = & \ID_f^*(1),\\
\frac{\delta}{\delta \Phi_j} \int_B fg & = & \ID_f^*(g) + (-1)^{|f||g|}
      \ID_g^*(f).
\label{eqn:frch}
\end{eqnarray}

We use (\ref{eqn:frch}) to determine when two functionals are in involution.
Namely, $\{{\eH}_n,{\eH}_m\}=0$ if and only if
\begin{equation}
\sum_{k=1}^{2}\left(
\left(\ID_{S_k}^j\right)^*\left(\frac{\delta {\eH}_m}
{\delta \Phi_k}\right)
+(-1)^{|S_k||\frac{\delta}{\delta\Phi_k}{\eH}_m|}
\left(\ID_{\frac{\delta}{\delta\Phi_k}{\eH}_m}^j\right)^*\left(S_k\right)
\right)=0
\end{equation}
for $j=1,2$, with $S_k\equiv {\eD}_{ik}\frac{\delta}{\delta \Phi_j}
{\eH}_n$. A program written in Mathematica \cite{mathematica} was used to
determine if this criterion is satisfied. The reason for working with
Fr\'{e}chet derivatives rather than with integrals is because we find
integration
to be more difficult to implement on the computer.

After a systematic search, we find the following three\footnote{
We explicitly ignore the three solutions corresponding to the $N=2$ super
KdV hierarchies \cite{laberge,labelle},
whose second Hamiltonian structure is the
$N=2$ Virasoro subalgebra of $N=2$ super $W_4$.}
systems of Hamiltonian
functionals mutually in involution:
\begin{eqnarray}
{\eH}^{-}_2 & = & \int_B P\left(a_1\Phi_2 + a_2 \Phi_1^2\right)
\\
{\eH}_3^{+} & = & \int_B Q\left( b_1\Phi_3+ b_2\Phi_1^3 +
	  b_3 \Phi_1\Phi_2 + b_4
	  \Phi_1(D_1D_2\Phi_1)
	  \right)
\\
{\eH}_4^{-} & = & \int_B P\left( c_1\Phi_2^{2}+ c_2\Phi_1^4 +
    c_3 \Phi_1\Phi_3 +  c_4 \Phi_1(D_1D_2\Phi_2) +
    c_5\Phi_1(\partial^2\Phi_1) \right. \nonumber\\
  & + & c_6 \Phi_1^2\Phi_2
     \left. + c_7 \Phi_1^2(D_1D_2\Phi_1)
	 \right)
\\
 {\eH}_5^{+} & = & \int_B Q\left( d_1\Phi^5 + d_2\Phi_2\Phi_3 +
  d_3 \Phi_2(D_1D_2\Phi_2) +
 d_4\Phi_2(\partial^2\Phi_1)\right.
  + d_5\Phi_2^2\Phi_1 \nonumber\\
  & + & d_6 \Phi_3(D_1D_2\Phi_1) +
  d_7\Phi_1(\partial^2D_1D_2\Phi_1) +
    d_8 \Phi_1^2\Phi_3 +
    d_9 \Phi_1^2(D_1D_2\Phi_2)\nonumber\\
  & + & d_{10}\Phi_1^2(\partial^2\Phi_1)
    + d_{11}\Phi_1^3\Phi_2
     + d_{12}\Phi_1^3(D_1D_2\Phi_1) +
    d_{13} \Phi_1(D_1D_2\Phi_1)^2  \nonumber\\
   & +  & \left. d_{14}
 \Phi_1\Phi_2(D_1D_2\Phi_1)
 \right)
\\
{\eH}_6^{-} & = & \int_B P\left( e_1 \Phi_2^3 + e_2 \Phi_3^2
    + e_3 \Phi_1^6 + e_4 (D_1D_2\Phi_2)^2
    + e_5 (D_1D_2\Phi_1)^3 \right.
 +  e_6 (\partial D_1D_2\Phi_1)^2 \nonumber\\
& + & e_7 \Phi_2(D_1D_2\Phi_3) + e_8 \Phi_2(\partial^2 D_1D_2\Phi_1)
    + e_9 \Phi_1^2\Phi_2^2 + e_{10} \Phi_2^2(D_1D_2\Phi_1) \nonumber\\
 & + & e_{11} \Phi_3(\partial^2\Phi_1)
 +   e_{12} \Phi_1^2(D_1D_2\Phi_3)
 + e_{13} \Phi_1^2(D_1D_2\Phi_1)^2 + e_{14} \Phi_1^2(\partial \Phi_1)^2
 \nonumber\\
 & + & e_{15} \Phi_1^2(\partial^2 \Phi_2)
+ e_{16} \Phi_1^2(\partial^2 D_1D_2 \Phi_1)
 +  e_{17}\Phi_1^3\Phi_3
 + e_{18} \Phi_1^3(D_1D_2\Phi_2)
  +  e_{19} \Phi_1^4\Phi_2 \nonumber\\
& + & e_{20} \Phi_1^4(D_1D_2\Phi_1)
 +  e_{21}\Phi_2(D_1D_2\Phi_1)^2
 +  e_{22} \Phi_2(\partial \Phi_1)^2
 +  e_{23} \Phi_1\Phi_2\Phi_3 \nonumber\\
& + & e_{24}\Phi_1\Phi_2(D_1D_2\Phi_2)
 + e_{25} \Phi_2(D_1\Phi_1)(\partial D_2\Phi_1) \nonumber\\
 & + & e_{26} \Phi_2(D_1\Phi_1)(\partial D_1\Phi_1)
 +  e_{27} \Phi_1\Phi_3(D_1D_2\Phi)
 + \left.  e_{28} \Phi_1^2\Phi_2(D_1D_2\Phi_1)\right).
\end{eqnarray}

The values of the coefficients such that the involution property is
satisfied are given in tables (1) - (4). From this, we postulate
the existence of three integrable hierarchies. The
flow equations
\begin{equation}
\frac{\partial \Phi_i}{\partial t} = {\eD}_{ij}\frac{\delta{\eH}_2^{-}}
{\delta \Phi_j}
\label{eqn:n2b}
\end{equation}
with $(a_1,a_2)=(1,\frac{7}{9}),(1,-\frac{2}{9}),(1,-\frac{5}{9})$ are the
three
$N=2$ supersymmetric generalized KdV equations associated with $N=2$
super $W_4$.
For two of the hierarchies,
non-trivial ${\eH}_n$ exist for every positive integer degree $n$, whereas
for the third hierarchy ${\eH}_{4k}$ is missing for $k=1$ and, we suspect,
for every integer $k$.
We have checked by hand that ${\eH}_3^{+}$ is indeed a
conservation law for the equation
(\ref{eqn:n2b}).
The existence of so many conservation
laws in involution can be regarded as very strong evidence that
three integrable hierarchies exist.

\begin{center}
\begin{tabular}{l|l|l|l|l} \hline\hline
$a_2/a_1$ & $b_1$ & $b_2$ & $b_3$ & $b_4$ \\ \hline
$\smallfrac{7}{9}$ & $-\smallfrac{10}{7}$ & $\smallfrac{20}{63}$ & $1$ &
$\smallfrac{5}{6}$
\\ \hline
$-\smallfrac{2}{9}$ & $20$ & $-\smallfrac{5}{18}$  & $1$ & $\smallfrac{5}{6}$
\\ \hline
$-\smallfrac{5}{9}$ & $\smallfrac{10}{3}$ & $-\smallfrac{10}{27}$ & $1$ &
$-\smallfrac{5}{6}$
\\ \hline\hline
\end{tabular}

\vspace{10pt}

Table 1. Values of $a_i$ and $b_i$ such that $\{{\eH}_2^{-},{\eH}_3^{+}\}=0$.
\end{center}
\begin{center}
\begin{tabular}{l|l|l|l|l|l|l|l} \hline\hline
$a_2/a_1$ & $c_1$ & $c_2$ & $c_3$ & $c_4$ & $c_5$ & $c_6$ & $c_7$\\ \hline
$\frac{7}{9}$ & $-\frac{1}{8}$ & $-\frac{49}{648}$ & $1$ & $-\frac{8}{15}$
      & $\frac{7}{18}$ & $-\frac{71}{180}$ & $-\frac{77}{216}$\\ \hline
$-\frac{5}{9}$ & $-\frac{1}{8}$ & $-\frac{49}{648}$ & $1$
      & $\frac{7}{15}$
      & $\frac{7}{18}$ & $\frac{49}{180}$ & $-\frac{77}{216}$\\ \hline\hline
\end{tabular}

\vspace{10pt}
Table 2. Values of $a_i$ and $c_i$ such that $\{{\eH}_2^{-},{\eH}_4^{-}\}=0$.
\end{center}

\begin{center}
\begin{tabular}{l|l|l|l|l|l|l|l|l|l} \hline\hline
$a_2/a_1$ & $d_1$ & $d_2$ & $d_3$ & $d_4$ &
$d_5$
& $d_6$ & $d_7$ & $d_8$ & $d_9$
\\ \hline
$\frac{7}{9}$ & $-\frac{46}{405}$ & $1$ & $-\frac{9}{20}$ & $\frac{3}{2}$
   & $-\frac{7}{10}$ & $\frac{10}{3}$ & $1$ & $\frac{25}{9}$
   & $-\frac{5}{8}$ \\ \hline
$-\frac{5}{9}$ & $\frac{89}{2430}$ & $1$ & $\frac{7}{40}$ & $\frac{1}{2}$
   & $\frac{2}{15}$ & $-\frac{5}{3}$ & $-\frac{7}{18}$ & $-\frac{5}{9}$
   & $-\frac{17}{72}$ \\ \hline
$-\frac{2}{9}$ & $-\frac{37}{12960}$ & $1$ & $\frac{7}{40}$ & $\frac{1}{12}$
   & $-\frac{3}{40}$ & $-\frac{5}{3}$ & $\frac{7}{72}$ & $\frac{5}{18}$
   & $-\frac{13}{144}$ \\ \hline\hline
\end{tabular}
\end{center}

\begin{center}
\begin{tabular}{l|l|l|l|l|l} \hline\hline
$a_2/a_1$ & $d_{10}$
& $d_{11}$ & $d_{12}$ & $d_{13}$ & $d_{14}$ \\ \hline
$\frac{7}{9}$  & $\frac{11}{12}$ & $-\frac{5}{6}$
   & $-\frac{89}{108}$ & $-\frac{11}{18}$ & $-\frac{23}{12}$\\ \hline
$-\frac{5}{9}$ & $-\frac{29}{72}$ & $-\frac{1}{6}$
   & $\frac{289}{972}$ & $\frac{19}{108}$ & $-\frac{13}{36}$\\ \hline
$-\frac{2}{9}$ & $-\frac{11}{108}$ & $\frac{1}{72}$
 & $\frac{61}{3888}$ & $-\frac{1}{8}$ & $\frac{1}{8}$\\ \hline\hline
\end{tabular}

\vspace{10pt}

Table 3. Values of $s_i$ and $d_i$ such that $\{{\eH}_2^{-},{\eH}_5^{+}\}=0$.
\end{center}

\begin{center}
\begin{tabular}{l|l|l|l|l|l|l|l|l|l|l} \hline\hline
$a_2/a_1$ & $e_1$ & $e_2$ & $e_3$ & $e_4$ & $e_5$
& $e_6$ & $e_7$ & $e_8$ & $e_9$ & $e_{10}$\\ \hline
$ \frac{7}{9}$ & $1$ & $6$ & $\frac{397}{729}$ & $\frac{207}{50}$
& $\frac{179}{54}$ & $\frac{15}{2}$ & $-\frac{72}{5}$ & $-12$
& $\frac{1151}{150}$
& $\frac{11}{2}$ \\ \hline
$-\frac{2}{9}$ & $1$ & $24$ & $-\frac{8}{729}$ & $-\frac{21}{25}$
& $-\frac{14}{27}$ & $-\frac{4}{3}$ & $\frac{72}{5}$ & $-4$ & $\frac{22}{75}$
& $-\frac{7}{2}$ \\ \hline
$ -\frac{5}{9}$ & $0$ & $1$ & $\frac{1}{81}$ & $\frac{9}{100}$
& $\frac{1}{12}$ & $\frac{1}{4}$ & $\frac{3}{5}$ & $\frac{3}{10}$
& $\frac{9}{100}$
& $0$ \\ \hline\hline
\end{tabular}
\end{center}

\begin{center}
\begin{tabular}{l|l|l|l|l|l|l|l|l|l|l} \hline\hline
$a_2/a_1$ & $e_{11}$ & $e_{12}$ & $e_{13}$  & $e_{14}$ &
$e_{15}$ & $e_{16}$ & $e_{17}$ & $e_{18}$ & $e_{19}$
& $e_{20}$  \\ \hline
$\frac{7}{9}$ & $30$ & $-\frac{23}{2}$ & $\frac{841}{108}$
& $\frac{827}{36}$
& $-\frac{33}{2}$  & $-\frac{23}{2}$ & $-20$
& $\frac{893}{180}$
& $\frac{139}{27}$ & $\frac{1807}{324}$
\\ \hline
$-\frac{2}{9}$ & $0$ & $-8$ & $-\frac{8}{27}$ & $\frac{32}{27}$
& $\frac{2}{3}$  & $\frac{2}{9}$ & $0$ & $-\frac{16}{45}$
& $\frac{4}{27}$ & $\frac{4}{81}$
\\ \hline
$-\frac{5}{9}$ & $1$ & $-\frac{7}{12}$ & $\frac{5}{24}$
& $\frac{59}{72}$
& $\frac{2}{5}$ & $-\frac{5}{12}$ & $-\frac{2}{9}$
& $-\frac{1}{8}$
& $-\frac{1}{15}$ & $\frac{11}{72}$
\\ \hline\hline
\end{tabular}
\end{center}

\begin{center}
\begin{tabular}{l|l|l|l|l|l|l|l|l} \hline\hline
$a_2/a_1$ & $e_{21}$ & $e_{22}$ & $e_{23}$ & $e_{24}$
& $e_{25}$ & $e_{26}$ & $e_{27}$ & $e_{28}$  \\ \hline
$ \frac{7}{9}$
& $\frac{191}{15}$ & $\frac{112}{5}$
& $-\frac{102}{5}$ & $\frac{237}{25}$ & $0$ & $-\frac{64}{5}$
& $-47$ & $\frac{4021}{180}$\\ \hline
$-\frac{5}{9}$
& $\frac{92}{15}$ & $\frac{4}{5}$
& $-\frac{48}{5}$ & $\frac{3}{25}$ & $0$ & $-\frac{28}{5}$
& $16$ & $-\frac{32}{45}$\\ \hline
$ -\frac{2}{9}$
& $-\frac{3}{20}$ & $-\frac{9}{20}$
& $\frac{3}{5}$
& $\frac{9}{50}$ & $0$ & $\frac{3}{10}$ & $-\frac{1}{2}$
& $-\frac{13}{40}$\\ \hline\hline
\end{tabular}

\vspace{10pt}

Table 4. Values of $a_i$ and $e_i$ such that $\{{\eH}_2^{-},{\eH}_6^{-}\}=0$.
\end{center}

\section{Conclusion}

In this paper, we have explicitly constructed the classical
$N=2$ super $W_4$ algebra. From a successful search for families of Hamiltonian
functionals in involution, we have argued for the existence of three
integrable hierarchies (invariant under space translations)
for which the classical $N=2$ super $W_4$ algebra furnish the (``second'')
Hamiltonian structure. We have not proved integrability of these
hierarchies, of course. We expect this to be a difficult task, given that
even in the case of the $N=2$ super Virasoro ($W_2$) algebra, integrability of
one of the hierarchies \cite{labelle} has yet to be
proven.  Nevertheless, we think that the strong evidence for the
existence of three integrable hierarchies  in each of the cases of the $N=2$
$W_2$, $W_3$ \cite{yung} and $W_4$ algebras suggest that this is the case
for all the $N=2$ super $W_n$ algebras of \cite{figueroa,inami,huitu}.

\vskip .2in
\noindent
{\Large \bf Acknowledgements} \newline
C.M.Y. thanks I.N.McArthur for helpful conversations. The authors acknowledge
th support of the Australian Research Council.

\vskip .2 in
\noindent
{\Large \bf Appendix} \newline
In this appendix, we present the explicit form for the nonlinear terms
of the matrix Hamiltonian operator ${\eD}_{ij}$ defining the Poisson
bracket (\ref{eqn:pb}). The diagonal terms ${\eD}_{22}$ and ${\eD}_{33}$
are written in manifestly anti-selfadjoint form for compactness. This form,
${\eD}_{ii}={\eO}_i - {\eO}_i^*$, is of
course non-unique.
\begin{eqnarray}
{\eO}_2 & = & -\smallfrac{5}{3}\partial^3 D_1D_2 -
   \smallfrac{10}{9} \Phi_1 \partial^3
 + \smallfrac{5}{3}(D_i\Phi_1)\partial^2D_i -\smallfrac{10}{9}(\partial \Phi_1)
    \partial^2
   + \smallfrac{10}{9}(D_1D_2\Phi_1)\partial D_1D_2 \nonumber\\
   &  - & \smallfrac{5}{27}\Phi_1^2
    \partial D_1D_2 - \smallfrac{7}{3} \Phi_2 \partial D_1D_2
    +\smallfrac{10}{9}(\partial D_i\Phi_1)\partial D_i
    + \smallfrac{115}{54}\Phi_1(D_1D_2\Phi_1)\partial\nonumber\\
   & - & \smallfrac{7}{15}\epsilon^{ij}(D_i\Phi_2)\partial D_j
   +\smallfrac{65}{54}\epsilon^{ij}\Phi_1(D_i\Phi_1)\partial D_j
   -\smallfrac{7}{15}(\partial \Phi_2)D_1D_2
   -\smallfrac{25}{18}\Phi_1(\partial\Phi_1)D_1D_2\nonumber\\
   & - & \smallfrac{49}{45}\Phi_1\Phi_2\partial
 -\smallfrac{10}{81}\Phi_1^3\partial
    -\smallfrac{5}{12}(D_1D_2\Phi_1)(D_i\Phi_1)D_i
   +  \smallfrac{5}{81}\Phi_1^2
    (D_i\Phi_1)D_i \nonumber\\
& - & 7\Phi_1(D_i\Phi_2)D_i + 7\Phi_2(D_i\Phi_1)D_i
   -  \smallfrac{5}{108}\epsilon^{ij}(\partial \Phi_1)(D_i\Phi_1)D_j
\nonumber\\
&  +  & 6 \Phi_3\partial - (D_i\Phi_3)D_i \nonumber
\end{eqnarray}

\begin{eqnarray}
{\eD}_{32} & = &
-\smallfrac{2}{5}(\partial D_1D_2\Phi_3)
+\smallfrac{3}{25}(\partial^3\Phi_2)
+\smallfrac{3}{5}\Phi_2(\partial \Phi_2)
-\smallfrac{8}{25}\Phi_2(\partial D_1D_2\Phi_1)\nonumber\\
& - & \smallfrac{38}{45}\Phi_1(\partial\Phi_3)
-\smallfrac{1}{25}\Phi_1(\partial D_1D_2\Phi_2)
+\smallfrac{2}{25}\Phi_1^2(\partial\Phi_2)
-\smallfrac{3}{25}(D_1D_2\Phi_2)(\partial \Phi_1)\nonumber\\
& - & \smallfrac{4}{5}(D_1D_2\Phi_3)\partial
-\smallfrac{8}{25}(D_1D_2\Phi_1)(\partial\Phi_2)
+\smallfrac{27}{50}(\partial^2\Phi_2)\partial
+\smallfrac{7}{25}\epsilon^{ij}(D_i\Phi_2)(\partial D_j\Phi_1)\nonumber\\
& - & \smallfrac{7}{9}(D_i\Phi_1)(D_i\Phi_3)
+\smallfrac{9}{50}\epsilon^{ij}(D_i\Phi_1)(\partial D_j\Phi_2)
-\smallfrac{4}{5}\epsilon^{ij}(\partial D_i\Phi_3)D_j \nonumber\\
& - &\smallfrac{9}{100}(\partial^2D_1D_2\Phi_2)D_i
 +\smallfrac{16}{75}\Phi_1\Phi_2(\partial\Phi_1)
-\smallfrac{16}{25}\Phi_2(D_1D_2\Phi_1)\partial \nonumber\\
& - &\smallfrac{3}{20}\Phi_2(D_i\Phi_2)D_i
-\smallfrac{1}{25}\epsilon^{ij}\Phi_2(\partial D_i\Phi_1)D_j
-\smallfrac{44}{15}\Phi_1\Phi_3\partial \nonumber\\
& + &2\Phi_3(D_i\Phi_1)D_i
-\smallfrac{3}{25}\Phi_1(D_1D_2\Phi_2)\partial
-\smallfrac{8}{45}\Phi_1(D_i\Phi_3)D_i  \nonumber\\
& + & \smallfrac{3}{100}\epsilon^{ij}\Phi_1(\partial D_i\Phi_2)D_j
+\smallfrac{14}{75}\Phi_1^2\Phi_2\partial
-\smallfrac{1}{50}\Phi_1^2(D_i\Phi_2)D_i \nonumber\\
& + &\smallfrac{3}{40}(D_1D_2\Phi_2)(D_i\Phi_1)D_i
+\smallfrac{3}{20}(D_1D_2\Phi_1)(D_i\Phi_2)D_i
-\smallfrac{3}{200}\epsilon^{ij}(\partial \Phi_2)(D_i\Phi_1)D_j \nonumber\\
& + &\smallfrac{9}{10}(\partial\Phi_2)\partial^2
-\smallfrac{28}{15}(\partial\Phi_3)D_1D_2
+ \smallfrac{3}{50}\epsilon^{ij}(\partial\Phi_1)(D_i\Phi_2)D_j \nonumber\\
& - &\smallfrac{3}{50}(\partial D_1D_2\Phi_2)D_1D_2
+\smallfrac{87}{200}(D_1\Phi_2)(D_2\Phi_1)\partial
+\smallfrac{87}{200}(D_1\Phi_1)(D_2\Phi_2)\partial  \nonumber\\
& - &\smallfrac{28}{15}\epsilon^{ij}(D_i\Phi_3)\partial D_j
-\smallfrac{9}{25}(\partial D_1\Phi_2)\partial D_i
-\smallfrac{4}{75}\Phi_1\Phi_2(D_i\Phi_1)D_i \nonumber\\
& + & \smallfrac{1}{25}\Phi_2(\partial\Phi_1)D_1D_2
 -  \smallfrac{3}{50}\epsilon^{ij}\Phi_2(D_i\Phi_1)\partial D_j
+\smallfrac{3}{5}\Phi_2\partial^3
-\smallfrac{28}{5}\Phi_3\partial D_1D_2 \nonumber\\
& + & \smallfrac{1}{25}\Phi_1(\partial \Phi_2)D_1D_2
+  \smallfrac{9}{100}\epsilon^{ij}\Phi_1(D_i\Phi_2)\partial D_j
-\smallfrac{9}{50}(D_1D_2\Phi_2)\partial D_1D_2\nonumber\\
& - & \smallfrac{1}{40}(D_i\Phi_2)(D_i\Phi_1)D_1D_2
-\smallfrac{9}{20}(D_i\Phi_2)\partial^2D_i
+\smallfrac{3}{25}\Phi_1\Phi_2\partial D_1D_2 \nonumber
\end{eqnarray}
\begin{eqnarray}
{\eO}_3 & = & -\smallfrac{1}{810}\Phi_1^5\partial   +
  \smallfrac{4}{25}(\partial^2\Phi_3)\partial  -
  \smallfrac{51}{500}(\partial^2D_1D_2\Phi_2)\partial  -
  \smallfrac{17}{80}(\partial^4\Phi_1)\partial -
  \smallfrac{1}{50}(\partial^2D_i\Phi_3)D_i \nonumber\\
& - &
  \smallfrac{17}{500}\epsilon^{ij}(\partial^3D_i\Phi_2)D_j +
  \smallfrac{17}{480}(\partial^4D_i\Phi_1)D_i +
  \smallfrac{2}{5}\Phi_2 \Phi_3\partial   -
  \smallfrac{11}{225}\Phi_3 \Phi_1^2\partial -
  \smallfrac{43}{2250}\Phi_2 \Phi_1^3\partial \nonumber\\
& + &
  \smallfrac{11}{100}\Phi_2(D_1D_2\Phi_2)\partial   -
  \smallfrac{11}{30}\Phi_2(\partial^2\Phi_1)\partial   +
  \smallfrac{1}{10}\Phi_2(D_i\Phi_3)D_i -
  \smallfrac{1}{400}\epsilon^{ij}(\Phi_2 (\partial D_i\Phi_2)D_j  \nonumber\\
& + &
  \smallfrac{97}{1000}\Phi_2 (\partial^2D_i\Phi_1)D_i -
  \smallfrac{21}{200}\Phi_2^2 \Phi_1\partial   +
  \smallfrac{21}{400}\Phi_2^2 (D_i\Phi_1)D_i
  +
  \smallfrac{53}{150}\Phi_3 (D_1D_2\Phi_1)\partial \nonumber\\
& - &\smallfrac{1}{4}\Phi_3 (D_i\Phi_2)D_i +
  \smallfrac{37}{600}\epsilon^{ij}\Phi_3 (\partial D_i\Phi_1)D_j +
  \smallfrac{3}{100}\Phi_1 (D_1D_2\Phi_3)\partial \nonumber\\
& - &\smallfrac{103}{480} \Phi_1 (D_1D_2\Phi_1)^2\partial
 + \smallfrac{809}{4320} \Phi_1 (\partial \Phi_1)^2\partial   +
  \smallfrac{109}{500} \Phi_1 (\partial^2\Phi_2)\partial \nonumber\\
& - &\smallfrac{13}{60} \Phi_1 (\partial^2D_1D_2\Phi_1)\partial   -
  \smallfrac{4}{75}\epsilon^{ij} \Phi_1 (\partial D_i\Phi_3)D_j
  -\smallfrac{1}{25}\Phi_1 (\partial^2D_i\Phi_2)D_i \nonumber\\
& - &\smallfrac{137}{480}\epsilon^{ij} \Phi_1 (\partial^3D_i\Phi_1)D_j -
  \smallfrac{481}{4500} \Phi_1^2 (D_1D_2\Phi_2)\partial +
  \smallfrac{1}{225}\Phi_1^2 (D_i\Phi_3)D_i \nonumber\\
& - &\smallfrac{1211}{12000}\epsilon^{ij} \Phi_1^2 (\partial D_i\Phi_2)D_j +
  \smallfrac{181}{2160} \Phi_1^2 (\partial^2D_i\Phi_1)D_i +
  \smallfrac{59}{2430} \Phi_1^3 (D_1D_2\Phi_1)\partial
   \nonumber\\
& + &\smallfrac{1}{4860}\epsilon^{ij}\Phi_1^3 (\partial D_i\Phi_1)D_j
   +\smallfrac{1}{1620} \Phi_1^4 (D_i\Phi_1)D_i
 - \smallfrac{19}{750} (D_1D_2\Phi_2) (D_1D_2\Phi_1)\partial
  \nonumber\\
& - &
  \smallfrac{1}{80} (D_1D_2\Phi_2) (D_i\Phi_2)D_i
 - \smallfrac{1}{1125}\Phi_1^3 (D_i\Phi_2)D_i
  +
  \smallfrac{5}{192} (D_1D_2\Phi_1)^2 (D_i\Phi_1)D_i
 \nonumber\\
  & + &\smallfrac{23}{600}\epsilon^{ij} (D_1D_2\Phi_2) (
\partial D_i\Phi_1)D_j -
  \smallfrac{1}{24} (D_1D_2\Phi_3) (D_i\Phi_1)D_i +
  \smallfrac{49}{144} (D_1D_2\Phi_1) (\partial^2\Phi_1)\partial \nonumber\\
  & - &\smallfrac{191}{3000}\epsilon^{ij} (D_1D_2\Phi_1) (
\partial D_i\Phi_2)D_j
  - \smallfrac{5}{288} (D_1D_2\Phi_1) (\partial^2D_i\Phi_1)D_i
  \nonumber\\
& - & \smallfrac{19}{8640} (\partial \Phi_1)^2 (D_i\Phi_1)D_i
 +\smallfrac{103}{360}\epsilon^{ij} (\partial^2\Phi_1) (
\partial D_i\Phi_1)D_j +
  \smallfrac{33}{100} \Phi_2 \Phi_1 (D_1D_2\Phi_1)\partial \nonumber\\
 & + &\smallfrac{713}{3600}\epsilon^{ij} \Phi_2 \Phi_1 (\partial D_i\Phi_1)D_j
  +\smallfrac{17}{1500} \Phi_2 (\Phi_1^2) (D_i\Phi_1)D_i -
  \smallfrac{3}{40} \Phi_2 (D_1D_2\Phi_1) (D_i\Phi_1)D_i \nonumber\\
& - &\smallfrac{23}{200} \Phi_2 (\partial \Phi_2)D_1D_2
  +
 \smallfrac{139}{720}\epsilon^{ij} \Phi_2 (\partial \Phi_1) (D_i\Phi_1)D_j -
  \smallfrac{173}{500} \Phi_2 (\partial \Phi_1)\partial^2
  -
  \smallfrac{7}{50} \Phi_3\partial^3
  \nonumber\\
  & + &\smallfrac{1}{100}\epsilon^{ij}\Phi_2 (D_i\Phi_2)\partial  D_j +
  \smallfrac{413}{1500} \Phi_2 (\partial D_i\Phi_1)\partial  D_i
  +
  \smallfrac{43}{750} \Phi_2 (\partial D_1D_2\Phi_1)D_1D_2
 \nonumber\\
 & + &\smallfrac{1}{90} \Phi_3 \Phi_1 (D_i\Phi_1)D_i -
  \smallfrac{19}{300} \Phi_3 (\partial \Phi_1)D_1D_2 +
  \smallfrac{17}{100}\epsilon^{ij} \Phi_3 (D_i\Phi_1)\partial  D_j
   - \smallfrac{23}{200} \Phi_2^2\partial  D_1D_2
  \nonumber\\
  & - &\smallfrac{23}{3600} \Phi_1 (D_1D_2\Phi_2) (D_i\Phi_1)D_i +
  \smallfrac{1}{80} \Phi_1 (D_1D_2\Phi_1) (D_i\Phi_2)D_i \nonumber\\
& - &\smallfrac{13}{80}\epsilon^{ij} \Phi_1 (D_1D_2\Phi_1) (
\partial D_i\Phi_1)D_j
  -\smallfrac{7}{60} \Phi_1 (\partial \Phi_3)D_1D_2 +
  \smallfrac{39}{500} \Phi_1 (\partial D_1D_2\Phi_2)D_1D_2 \nonumber\\
& + &\smallfrac{7}{24} \Phi_1 (\partial D_1D_2\Phi_1)\partial^2
   - \smallfrac{31}{120} \Phi_1 (\partial^3\Phi_1)D_1D_2 -
  \smallfrac{41}{250} \Phi_1 (\partial D_i\Phi_2)\partial  D_i \nonumber\\
& - &\smallfrac{211}{480}\epsilon^{ij} \Phi_1 (\partial^2D_i\Phi_1)\partial D_j
  -\smallfrac{13}{1620} \Phi_1^2 (D_1D_2\Phi_1) (D_i\Phi_1)D_i -
  \smallfrac{1817}{18000} \Phi_1^2 (\partial \Phi_2)D_1D_2 \nonumber\\
& + &\smallfrac{5}{96} \Phi_1^2 (\partial \Phi_1)\partial^2
   + \smallfrac{367}{4320} \Phi_1^2 (\partial D_1D_2\Phi_1)D_1D_2 -
  \smallfrac{83}{750}\epsilon^{ij} \Phi_1^2 (D_i\Phi_2)\partial  D_j
\nonumber\\
& + &\smallfrac{1}{50}  (D_1D_2\Phi_3)\partial  D_1D_2
  +\smallfrac{419}{4320} \Phi_1^2 (\partial D_i\Phi_1)\partial  D_i -
  \smallfrac{11}{1215} \Phi_1^3 (\partial \Phi_1)D_1D_2 \nonumber\\
& + &\smallfrac{17}{2430} \Phi_1^3 (D_i\Phi_1)\partial  D_j -
  \smallfrac{1}{72} \Phi_1^3\partial^3
   - \smallfrac{1}{810} \Phi_1^4\partial  D_1D_2  +
  \smallfrac{9}{125} (D_1D_2\Phi_2) (\partial \Phi_1)D_1D_2\nonumber\\
& + &\smallfrac{7}{250} (D_1D_2\Phi_2)\partial^3
 - \smallfrac{5}{32}\epsilon^{ij}(D_1D_2\Phi_1) (\partial \Phi_1)(D_i\Phi_1)D_j
\nonumber\\
& + &\smallfrac{59}{360} (D_1D_2\Phi_1) (\partial D_1D_2\Phi_1)D_1D_2
  +\smallfrac{17}{1500}\epsilon^{ij} (D_1D_2\Phi_1) (D_i\Phi_2)\partial  D_j
\nonumber\\
& - &\smallfrac{247}{720} (D_1D_2\Phi_1) (\partial D_i\Phi_1)\partial  D_i -
  \smallfrac{2}{45} (D_1D_2\Phi_1)^2\partial  D_1D_2       \nonumber\\
  & - & \smallfrac{2}{5} (\partial \Phi_1)^2\partial  D_1D_2 +
  \smallfrac{21}{250} (\partial^2\Phi_2)\partial  D_1D_2
+ \smallfrac{29}{160}\epsilon^{ij} (\partial^2\Phi_1) (D_i\Phi_1)
\partial  D_j   \nonumber\\
& + &\smallfrac{1}{6} (\partial^2\Phi_1)\partial^3  -
  \smallfrac{1}{20}(\partial^2D_1D_2\Phi_1)\partial  D_1D_2 -
  \smallfrac{17}{1500}\epsilon^{ij} (D_i\Phi_2) (\partial D_j\Phi_1)D_1D_2
  \nonumber\\
  &+ &\smallfrac{7}{100} (D_i\Phi_3)\partial^2 D_i +
  \smallfrac{9}{125}\epsilon^{ij} (D_i\Phi_1) (\partial D_j\Phi_2)D_1D_2 +
  \smallfrac{9}{160} (D_i\Phi_1) (\partial^2D_i\Phi_1)D_1D_2 \nonumber\\
& + &
  \smallfrac{21}{250}\epsilon^{ij} (\partial D_i\Phi_2)\partial^2 D_j -
  \smallfrac{1}{8}(\partial^2D_i\Phi_1)\partial^2 D_i
  -
  \smallfrac{7}{120} (D_i\Phi_3) (D_i\Phi_1)D_1D_2
  \nonumber\\
  & + &\smallfrac{1639}{9000} \Phi_2 \Phi_1 (\partial \Phi_1)D_1D_2 +
  \smallfrac{857}{4500}\epsilon^{ij} \Phi_2 \Phi_1 (D_i\Phi_1)\partial  D_j -
  \smallfrac{14}{125} \Phi_2 \Phi_1\partial^3 \nonumber\\
& - &\smallfrac{28}{1125} \Phi_2 \Phi_1^2\partial  D_1D_2
+\smallfrac{56}{375} \Phi_2 (D_1D_2\Phi_1)\partial  D_1D_2
  +\smallfrac{49}{250} \Phi_2 (D_i\Phi_1)\partial^2 D_i \nonumber\\
& - &\smallfrac{1}{50}\Phi_3 \Phi_1\partial  D_1D_2 -
\smallfrac{167}{1080} \Phi_1 (D_1D_2\Phi_1) (\partial \Phi_1)D_1D_2 \nonumber\\
& - &
\smallfrac{167}{1080}\epsilon^{ij} \Phi_1 (D_1D_2\Phi_1)
(D_i\Phi_1)\partial D_j +
  \smallfrac{11}{48} \Phi_1 (D_1D_2\Phi_1)\partial^3 \nonumber\\
 & + &\smallfrac{7}{432} \Phi_1 (\partial \Phi_1) (D_i\Phi_1)\partial  D_i
-\smallfrac{2}{5} \Phi_1 (\partial^2\Phi_1)\partial  D_1D_2 -
  \smallfrac{1}{72} \Phi_1 (D_i\Phi_2) (D_i\Phi_1)D_1D_2 \nonumber\\
& - &\smallfrac{7}{500} \Phi_1 (D_i\Phi_2)\partial^2 D_i
    +\smallfrac{2}{135} \Phi_1^2 (D_1D_2\Phi_1)\partial  D_1D_2 +
  \smallfrac{1}{48}  \Phi_1^2 (D_i\Phi_1)\partial^2 D_i \nonumber\\
& - &\smallfrac{13}{96} (D_1D_2\Phi_1) (D_i\Phi_1)\partial^2 D_i
 - \smallfrac{1}{216}(\partial \Phi_1) (D_1\Phi_1) (D_2\Phi_1)D_1D_2
\nonumber\\
& - &\smallfrac{7}{500}\epsilon^{ij} (D_i\Phi_2) (D_j\Phi_1)\partial  D_1D_2
+ \smallfrac{1}{240} (D_i\Phi_1) (\partial D_i\Phi_1)\partial D_1D_2\nonumber\\
& - &\smallfrac{7}{50} \Phi_2\partial^3 D_1D_2 -
  \smallfrac{5}{12} \Phi_1 (\partial \Phi_1)\partial^2 D_1D_2
  +
  \smallfrac{1}{96}\epsilon^{ij}\Phi_1 (\partial D_i\Phi_1)\partial^2 D_j
  \nonumber\\
& - &\smallfrac{1}{720}\Phi_1 (D_1\Phi_1) (D_2\Phi_1)\partial  D_1D_2 +
  \smallfrac{19}{144}\epsilon^{ij} \Phi_1 (D_i\Phi_1)\partial^3 D_j
  -
  \smallfrac{1}{72}\Phi_1^2\partial^3 D_1D_2
  \nonumber\\
  & - &\smallfrac{1}{40}\Phi_1\partial^5
  +
  \smallfrac{1}{12} (D_1D_2\Phi_1)\partial^3 D_1D_2 +
  \smallfrac{1}{16} (D_i\Phi_1)\partial^4 D_i -
  \smallfrac{1}{40}\partial^5 D_1D_2 \nonumber
\end{eqnarray}

\end{document}